\documentclass[aps, prd, twocolumn, superscriptaddress, showpacs, amsmath, amssymb, tightenlines]{revtex4-1}

\usepackage{graphicx} % Include figure files
\usepackage{dcolumn}  % Align table columns on decimal point
\usepackage{epstopdf}
\usepackage{bm}
\usepackage{nicefrac}

\usepackage[absolute]{textpos}

\graphicspath{{ps}}

\begin{document}

\title{First observation of \boldmath{$CP$} violation and improved measurement of the branching fraction and polarization of \boldmath{$B^0 \rightarrow D^{*+} D^{*-}$} decays}

\affiliation{Budker Institute of Nuclear Physics SB RAS and Novosibirsk State University, Novosibirsk 630090}
\affiliation{Faculty of Mathematics and Physics, Charles University, Prague}
\affiliation{University of Cincinnati, Cincinnati, Ohio 45221}
\affiliation{Gifu University, Gifu}
\affiliation{Hanyang University, Seoul}
\affiliation{University of Hawaii, Honolulu, Hawaii 96822}
\affiliation{High Energy Accelerator Research Organization (KEK), Tsukuba}
\affiliation{Indian Institute of Technology Guwahati, Guwahati}
\affiliation{Indian Institute of Technology Madras, Madras}
\affiliation{Institute of High Energy Physics, Chinese Academy of Sciences, Beijing}
\affiliation{Institute of High Energy Physics, Vienna}
\affiliation{Institute of High Energy Physics, Protvino}
\affiliation{Institute for Theoretical and Experimental Physics, Moscow}
\affiliation{J. Stefan Institute, Ljubljana}
\affiliation{Kanagawa University, Yokohama}
\affiliation{Institut f\"ur Experimentelle Kernphysik, Karlsruher Institut f\"ur Technologie, Karlsruhe}
\affiliation{Korea Institute of Science and Technology Information, Daejeon}
\affiliation{Korea University, Seoul}
\affiliation{Kyungpook National University, Taegu}
\affiliation{\'Ecole Polytechnique F\'ed\'erale de Lausanne (EPFL), Lausanne}
\affiliation{Faculty of Mathematics and Physics, University of Ljubljana, Ljubljana}
\affiliation{University of Maribor, Maribor}
\affiliation{Max-Planck-Institut f\"ur Physik, M\"unchen}
\affiliation{University of Melbourne, School of Physics, Victoria 3010}
\affiliation{Graduate School of Science, Nagoya University, Nagoya}
\affiliation{Kobayashi-Maskawa Institute, Nagoya University, Nagoya}
\affiliation{Nara Women's University, Nara}
\affiliation{National Central University, Chung-li}
\affiliation{National United University, Miao Li}
\affiliation{Department of Physics, National Taiwan University, Taipei}
\affiliation{H. Niewodniczanski Institute of Nuclear Physics, Krakow}
\affiliation{Nippon Dental University, Niigata}
\affiliation{Niigata University, Niigata}
\affiliation{University of Nova Gorica, Nova Gorica}
\affiliation{Osaka City University, Osaka}
\affiliation{Pacific Northwest National Laboratory, Richland, Washington 99352}
\affiliation{Panjab University, Chandigarh}
\affiliation{Seoul National University, Seoul}
\affiliation{Sungkyunkwan University, Suwon}
\affiliation{School of Physics, University of Sydney, NSW 2006}
\affiliation{Tata Institute of Fundamental Research, Mumbai}
\affiliation{Excellence Cluster Universe, Technische Universit\"at M\"unchen, Garching}
\affiliation{Tohoku Gakuin University, Tagajo}
\affiliation{Tohoku University, Sendai}
\affiliation{Department of Physics, University of Tokyo, Tokyo}
\affiliation{Tokyo Institute of Technology, Tokyo}
\affiliation{Tokyo Metropolitan University, Tokyo}
\affiliation{Tokyo University of Agriculture and Technology, Tokyo}
\affiliation{CNP, Virginia Polytechnic Institute and State University, Blacksburg, Virginia 24061}
\affiliation{Yamagata University, Yamagata}
\affiliation{Yonsei University, Seoul}

  \author{B.~Kronenbitter}\affiliation{Institut f\"ur Experimentelle Kernphysik, Karlsruher Institut f\"ur Technologie, Karlsruhe} % Karlsruhe
  \author{I.~Adachi}\affiliation{High Energy Accelerator Research Organization (KEK), Tsukuba} % KEK
  \author{H.~Aihara}\affiliation{Department of Physics, University of Tokyo, Tokyo} % Tokyo
  \author{K.~Arinstein}\affiliation{Budker Institute of Nuclear Physics SB RAS and Novosibirsk State University, Novosibirsk 630090} % BINP
  \author{D.~M.~Asner}\affiliation{Pacific Northwest National Laboratory, Richland, Washington 99352} % PNNL
  \author{T.~Aushev}\affiliation{Institute for Theoretical and Experimental Physics, Moscow} % ITEP
  \author{T.~Aziz}\affiliation{Tata Institute of Fundamental Research, Mumbai} % Tata
  \author{A.~M.~Bakich}\affiliation{School of Physics, University of Sydney, NSW 2006} % Sydney
  \author{M.~Barrett}\affiliation{University of Hawaii, Honolulu, Hawaii 96822} % Hawaii
  \author{K.~Belous}\affiliation{Institute of High Energy Physics, Protvino} % Protvino
% ...
% Karlsruhe
\collaboration{The Belle Collaboration}
\noaffiliation
%% end author list

\begin{abstract}
We report the measurement of the branching fraction, the polarization, and the parameters of the time-dependent $CP$ violation in $B^0 \rightarrow D^{*+} D^{*-}$ decays using a data sample of $772 \times 10^6 B \bar{B}$~pairs, collected at the $\Upsilon(4S)$ resonance with the Belle detector at the KEKB asymmetric-energy $e^+e^-$ collider. We obtain a branching fraction of $\mathcal{B} = (7.82 \pm 0.38 \pm 0.63) \times 10^{-4}$, a $CP$-odd fraction of $R_\perp = 0.138 \pm 0.024 \pm 0.006$ and, additionally, a fraction of the longitudinal component in the transversity base of $R_0 = 0.624 \pm 0.029 \pm 0.011$. The measured values of the parameters of the $CP$ violation are $S_{D^{*+}D^{*-}} = -0.79 \pm 0.13 \pm 0.03$ and $A_{D^{*+}D^{*-}} = 0.15 \pm 0.08 \pm 0.04$.
\end{abstract}

\pacs{11.30.Er, 12.15.Hh, 13.25.Hw}

\maketitle

%%%% >>>> keep the final version single-spaced

{\renewcommand{\thefootnote}{\fnsymbol{footnote}}}
\setcounter{footnote}{0}
\par
The analysis of the time-dependent decay rate of neutral $B$ mesons allows for the measurement of $CP$ violation in the $B$ system, which is related to the complex phase in the Cabibbo-Kobayashi-Maskawa (CKM) quark mixing matrix~\cite{C, KM}. The decays of a neutral $B$ meson to two charged charm mesons have been analyzed in numerous final states both by the Belle and the BaBar collaborations \cite{DC_BaBar, DC_Belle1, DC_Belle2}. The mixing-induced $CP$ violation in these decays is related to one of the angles of the CKM triangle $\phi_1 = \arg \left(- V_{cd}V^*_{cb}/V_{td}V^*_{tb} \right)$, which was precisely measured in $b \to (c\overline{c}) s$ decays~\cite{jpsiks,sin2phi1BaBar}.

\par
The time-dependent decay rate of a neutral $B$ meson, originating from the decay of an $\Upsilon (4S)$, to a $CP$-eigenstate is given by
\begin{multline}
\mathcal{P}_{B^0} \left(\Delta t \right) = \frac{1}{4 \tau_{B^0}} e^{{}^{-|\Delta t|}/{}_{\tau_{B^0}}} \\ \times \{ 1 + q \left[ S \sin \left( \Delta m  \Delta t \right) + A \cos \left( \Delta m \Delta t \right) \right] \},
\label{equ:dt_sig}
\end{multline}
where $q$ is either $+1$ or $-1$ when the non-signal $B$ meson decays as $B^0$ or $\bar{B}^0$ respectively, $\Delta t$ is the decay time difference of the two $B$ mesons, $\tau_{B^0}$ is the lifetime of the neutral $B$ meson, and $\Delta m$ is the mass difference of the two mass-eigenstates of the neutral $B$ system. 

\par
The decay $B^0 \rightarrow D^{*+} D^{*-}$ is the decay of a pseudo-scalar to two vector mesons and the final state is not a pure $CP$-eigenstate but a mixture of $CP$-even and $CP$-odd final states, depending on the relative angular momentum of the $D^*$ mesons~\cite{CC}. For an angular momentum of zero or two, the final state is $CP$-even while, for an angular momentum of one, it is $CP$-odd.
The dominant contribution to the decay $B^0 \rightarrow D^{*+} D^{*-}$ is the tree-level $b \to c \bar{c} d$ transition. Contributions from penguin diagrams are possible but strongly suppressed in the Standard Model \cite{Xing}. 

\par
The measurement is performed using the final Belle data sample consisting of an integrated luminosity of $711 \, \text{fb}^{-1}$ containing $(772 \pm 11) \times 10^6 B \bar{B}$ pairs, collected at the $\Upsilon(4S)$ resonance at the KEKB asymmetric-energy $e^+e^-$ collider~\cite{kekb}. The electron beam has an energy of $8$ GeV and the positron beam $3.5$ GeV, which leads to a boost of the $\Upsilon (4S)$ of $\beta \gamma = 0.425$ along the beam axis. Therefore, the neutral $B$ mesons have an average absolute flight length difference of about $200$ $\mathrm{\mu m}$, which allows for the determination of the decay time difference $\Delta t$. 

% For the determination of the reconstruction efficiency and the shape of the signal distributions, we generate multiple samples of simulated signal events. Therefore we first simulate the decays to the final state, using the software package EvtGen~\cite{evtgen} and then the interaction with the detector and its response, using GEANT3~\cite{GEANT}. The simulated signal events are overlaid by beam related background, which was recorded with a random trigger.

% {\it SVD2+SVD1:}
The Belle detector is a large-solid-angle magnetic spectrometer that consists of a silicon vertex detector (SVD), a 50-layer central drift chamber (CDC), an array of aerogel threshold Cherenkov counters (ACC), a barrel-like arrangement of time-of-flight scintillation counters, and an electromagnetic calorimeter composed of CsI(Tl) crystals (ECL) located inside a superconducting solenoid coil that provides a 1.5~T magnetic field.  An iron flux-return located outside the coil is instrumented to detect $K_L^0$ mesons and to identify muons. The detector is described in detail elsewhere~\cite{Belle}.
% {\bf SVD2+SVD1, up to experiment 37:}
Two inner detector configurations were used. A 2.0 cm beam-pipe and a 3-layer silicon vertex detector were used for the first sample of $152 \times 10^6 B\bar{B}$ pairs, while a 1.5 cm beam-pipe, a 4-layer silicon detector and a small-cell inner drift chamber were used to record the remaining $620 \times 10^6 B\bar{B}$ pairs~\cite{svd2}.
In 2011, the data set recorded with the second configuration of the SVD was reprocessed using new track finding algorithms. This significantly improved the track reconstruction efficiency, especially for low momentum tracks. Due to the presence of up to two charged slow pions in the final state and the high multiplicity of the decay, this leads to an increase in the reconstruction efficiency of $B^0 \rightarrow D^{*+} D^{*-}$ decays of $79 \%$ in comparison to the last Belle measurement, which used $657 \times 10^6 B \bar{B}$ pairs~\cite{DC_Belle2}.

\par
The analysis of $B^0 \rightarrow D^{*+} D^{*-}$ decays is performed by reconstructing charged $D^{*}$ mesons via $D^{*+} \to D^0 \pi^+$ and $D^{*+} \to D^+ \pi^0$.
Charged $D$ mesons are reconstructed in the modes $K^- \pi^+ \pi^+$, $K^0_S \pi^+$, $K^0_S \pi^+ \pi^0$, and $K^+ K^- \pi^+$, and neutral $D$ mesons in $K^- \pi^+$, $K^- \pi^+ \pi^0$, $K^- \pi^+ \pi^+ \pi^-$, $K^0_S \pi^+ \pi^-$, and $K^+ K^-$.
Neutral pions are reconstructed in the $\pi^0 \to \gamma \gamma$ decay mode from photons with an energy greater than $30$ MeV and are required to have an invariant mass that lies within $15 \, \mathrm{MeV/\it{c} \rm^2}$ of the nominal mass~\cite{PDG}; this corresponds to a width of $3.3 \, \sigma$. 
The neutral kaons, reconstructed in $K_S^0 \to \pi^+ \pi^-$ decays, must have an invariant mass in the range $470$ $\mathrm{MeV/\it{c} \rm^2} < m_{\pi\pi} < 520$ $\mathrm{MeV/ \it{c} \rm ^2}$ and fulfill the criteria described in Ref.~\cite{Ks}.
Charged tracks used for the reconstruction of $D$ mesons are required to have a distance to the interaction point along (perpendicular to) the beam direction of less than $4 ~ (2)$ cm. We separate charged kaons and pions based on the information from the ACC, the time-of-flight measurement, and the measurement of the energy loss in the CDC. The applied selection has an efficiency of $98 \%$ $(97 \%)$ and a misidentification rate of $9 \%$ $(18 \%)$ for pions (kaons). Additionally, an electron veto based on the measurement of the shower shape and the energy deposit in the ECL is applied. 
Pions that are used to form $D^*$ mesons are denoted as slow pions, due to their low momentum, and need not pass particle identification and the distance requirement because of their low spatial resolution and short range in the detector. 
\par
Final states containing more than one $D^{*+} \to D^+ \pi^0$ candidate or more than one $K^0_S$ candidate are discarded due to their high background level.
$D$ and $D^*$ candidates are selected based on their invariant mass and the mass difference between the $D^*$ and the $D$ meson candidate. The applied selection is decay mode dependent and has an efficiency of about $95 \%$ for the decays of $D$ mesons to two charged tracks and $85$ to $90 \%$ otherwise. These requirements as well as the selection criteria for kaons and pions are determined using Monte Carlo (MC) simulated events by optimizing a figure-of-merit $S / \sqrt{S + B}$, where $S$ is the number of signal events and $B$ is the number of background events.
Charged tracks originating from a $D$ meson decay are constrained to have a common origin. The slow pions are constrained to originate from the decay vertex of the $D^*$ mesons, which is obtained by projecting the $D$ momentum vectors to the beam intersection region. Two $D^*$ candidates are combined to form a $B^0$ candidate. Its decay vertex is determined by a kinematic fit of the tracks of the two $D$ mesons to a common vertex with the constraint that they originate from the beam interaction region.

\par
Two variables, the beam-constrained-mass $M_{\rm bc} = \sqrt{\nicefrac{s}{4} - \left( \vec{p}^{\; *}_{B} \right)^2}$ and the energy difference $\Delta E = E^{*}_{B} - \nicefrac{\sqrt{s}}{2}$, with $\nicefrac{\sqrt{s}}{2}$ being the beam energy in the center-of-mass system, $\vec{p}^{\; *}_{B}$ the momentum, and $E^{*}_{B}$ the energy of the fully reconstructed $B$ meson, also in the center-of-mass system, are used to discriminate between signal decays and the background. Events with $M_{\rm bc} < 5.23$ $\mathrm{GeV/\it{c} \rm^2}$ or $\left| \Delta E \right| > 0.14$ GeV are rejected. We also define a tighter selection region, referred to as the signal region, with $M_{\rm bc} > 5.27$ $\mathrm{GeV/\it{c} \rm^2}$ and $|\Delta E| < 0.04$ GeV. This region is used to optimize the selection criteria as described above.
In $32 \%$ of the signal events, there are multiple candidates. In the case of multiple candidates, the best candidate is selected based on the masses of the $D$ mesons and the mass differences between the $D^*$ and $D$ meson candidates.

\par
The selected sample consists of combinatorial background and signal. Not all signal events are completely correctly reconstructed: in some, a neutral pion is wrongly reconstructed or a charged kaon is misidentified. This component, denoted as cross-feed, is treated as signal and fitted with its own probability density function (PDF).

\par
\begin{figure}[ht]
 \includegraphics[width=.45\textwidth]{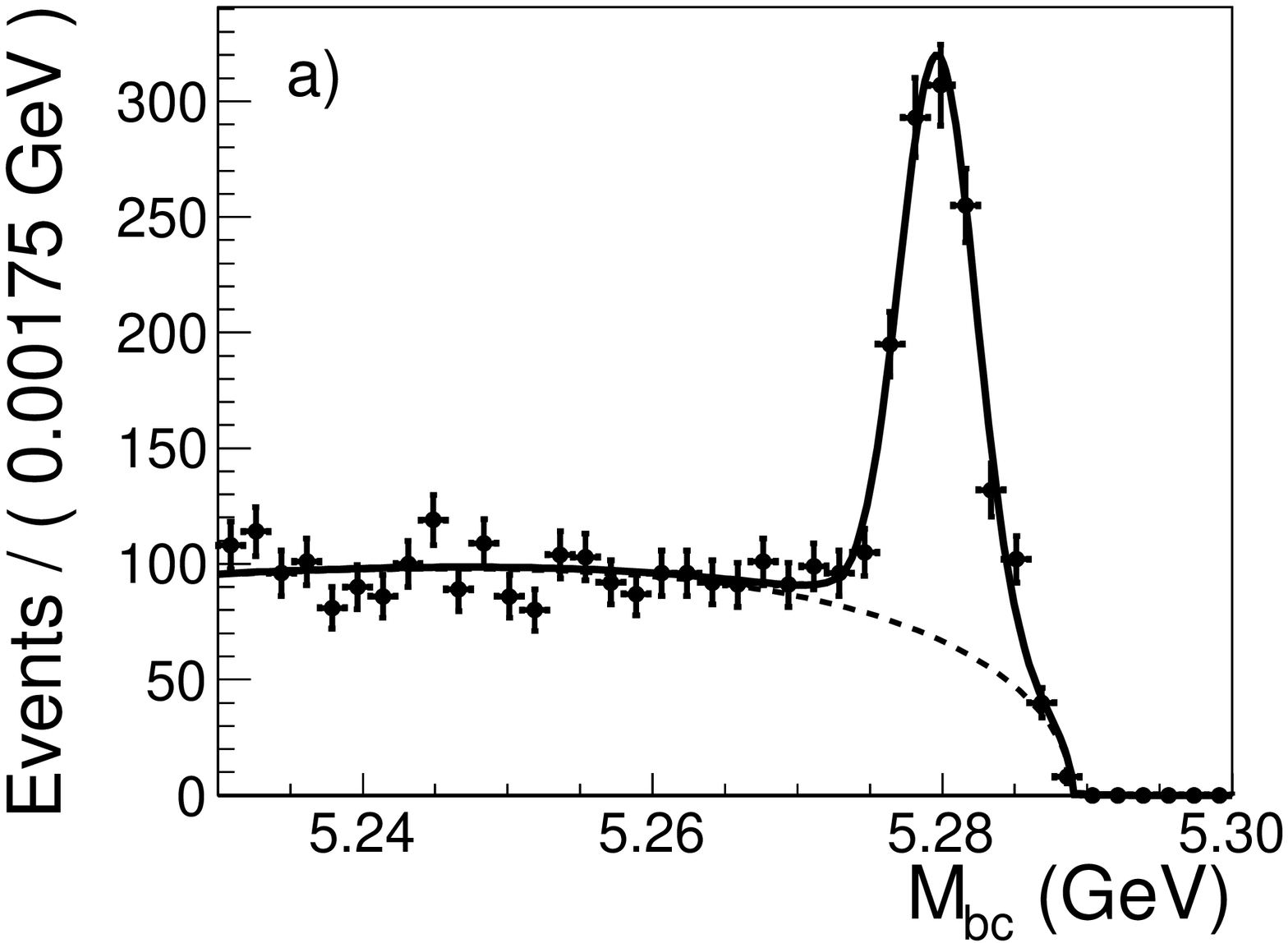} \\
 \includegraphics[width=.45\textwidth]{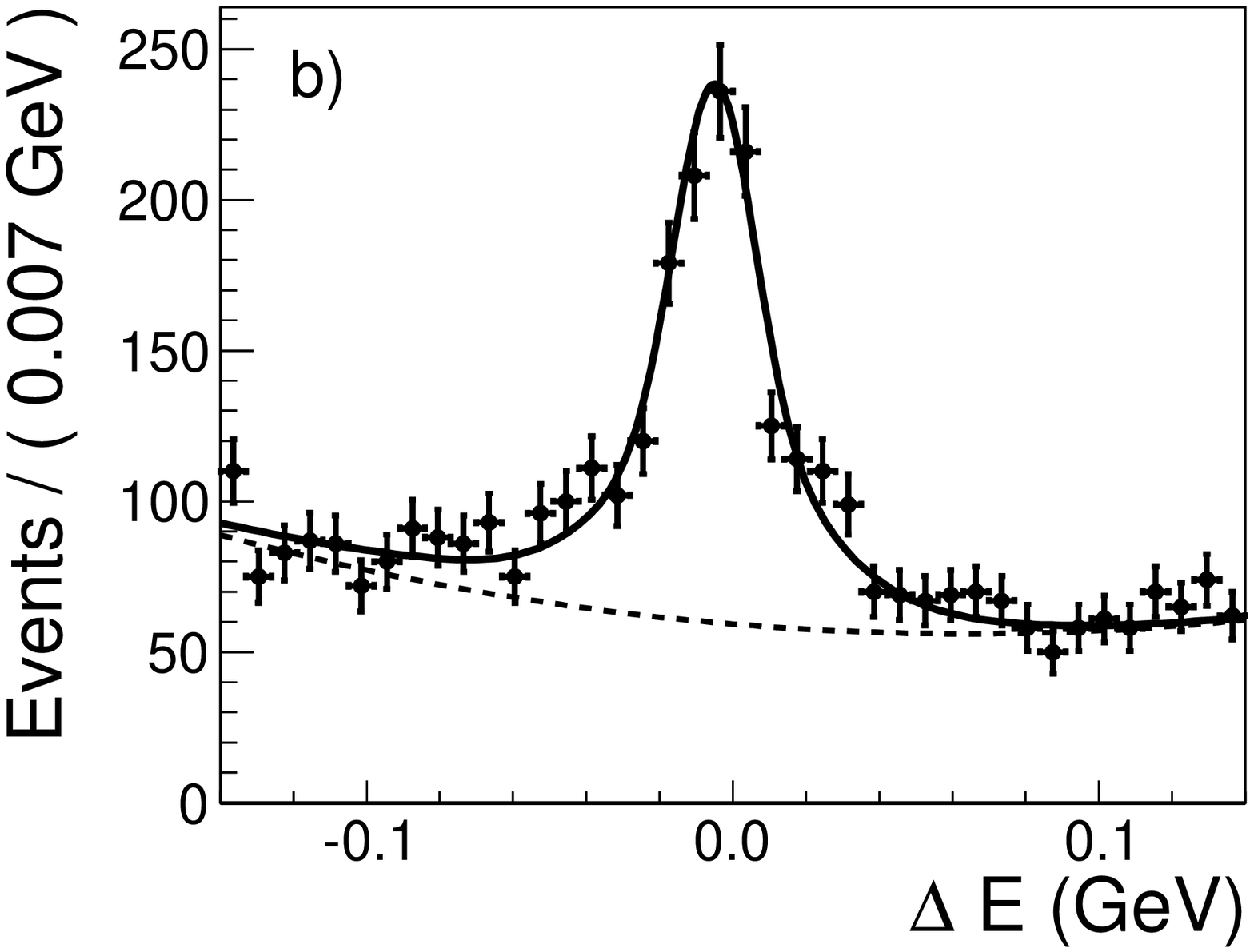}
 \caption{Fit projections and data points of $M_{\rm bc}$ (top) and $\Delta E$ (bottom). The solid line shows the total fitted distribution in the signal region of the other variable. The dashed line shows the fitted background distribution.}
\label{fig:mbcde}
\end{figure}

The number of reconstructed signal events, used for the calculation of the branching fraction, is determined by an extended, two-dimensional, unbinned maximum likelihood fit of the $M_{\rm bc}$ and $\Delta E$ distributions. The signal is described in $\Delta E$ by the sum of two Gaussians and a bifurcated Gaussian with a common mean, and in $M_{\rm bc}$ by an empirically determined parameterized signal shape introduced by the Crystal Ball collaboration~\cite{crb}. The fractions of the individual terms and the ratio of the widths are determined using simulated signal events, while the mean and a scale factor of the width, which is introduced to absorb the possible difference between data and simulation, are floated in the fit.
The cross-feed is described by an additional term consisting of the sum of a constant term and a Crystal Ball function in $\Delta E$, and a single Crystal Ball function in $M_{\rm bc}$. Its shape and fraction are determined using simulated signal events and fixed in the fit. The fraction of the cross-feed is found to be $11.6 \%$ of the total signal yield.
The combinatorial background is described by a second-order polynomial in $\Delta E$ and an empirically determined parametrized background shape introduced by the ARGUS collaboration~\cite{argus} in $M_{\rm bc}$.
The fitted number of signal events is $1225 \pm 59$. The fit projections and data distributions are shown in Fig.~\ref{fig:mbcde}. Together with the reconstruction efficiency, obtained using Monte Carlo simulated signal events, and the branching fractions of the decays of $D^*$ and $D$ mesons, taken from Ref.~\cite{PDG}, this gives a branching fraction of $\mathcal{B}(B^0 \rightarrow D^{*+} D^{*-}) = (7.82 \pm 0.38) \times 10^{-4}$. 

\par
The contributions to the systematic error of the branching fraction are listed in Table~\ref{tab:sys_br}. The dominant errors are the uncertainty of the reconstruction efficiency of charged tracks, neutral pions, and slow pions, and of the selection performed for the separation of charged kaons and pions (PID). These uncertainties have been estimated using high-statistics control samples.
The fit model error is determined by varying all parameters that are obtained from simulation and fixed in the fit by two standard deviations ($\sigma$) to account for the statistical uncertainties of the Monte Carlo samples and possible systematic differences between data and simulation. We repeat the fit for each parameter and add all deviations from the nominal fit result in quadrature, where the fraction of the cross-feed is varied by $20 \%$ of its value. The event reconstruction efficiency error includes the uncertainty of reconstruction efficiencies as well as uncertainties in efficiencies of selection criteria related to the masses of the $D$ and $D^*$ mesons. The uncertainty in polarization has a negligible effect upon the overall systematic error. The total error is calculated by adding all contributions in quadrature.

\par
In total, we measure a branching fraction of $\mathcal{B}_{(B^0 \rightarrow D^{*+} D^{*-})} = \left( 7.82 \pm 0.38 \pm 0.63 \right) \times 10^{-4},$ where the first error is statistical and the second is systematic. This is consistent with previous measurements made by the Belle, BaBar, and CLEO collaborations, which were performed using data samples having sizes of $1-30 \%$ the size of our sample~\cite{BRF_Belle,BRF_BaBar,BRF_CLEO}.

\begin{table}[ht]
\caption{Systematic errors of the branching fraction.}
\label{tab:sys_br}
\begin{tabular}
 {@{\hspace{0.5cm}}l@{\hspace{0.10cm}} @{\hspace{0.10cm}}c@{\hspace{0.5cm}}}
\hline \hline
Source & Systematic error (\%) \\
\hline
Charged track reconstruction & $\pm 1.7$ \\
$K_S$ reconstruction & $\pm 0.8$ \\
$\pi^0$ reconstruction & $\pm 3.0$ \\
Slow pion reconstruction & $\pm 3.2$ \\
PID selection efficiency & $\pm 5.0$ \\
$N_{B \bar{B}}$ & $\pm 1.4$ \\
Fit model & $\pm 2.1$ \\
$D$ and $D^*$ branching fractions & $\pm 3.1$ \\
Event reconstruction & $\pm 0.8$ \\
\hline
Total & $\pm 8.1$ \\
\hline \hline
\end{tabular}
\end{table}

\par
For the measurement of the parameters of $CP$ violation, additional requirements are made as described in Ref.~\cite{jpsiks}. 
The parameters $S$ and $A$ are determined with a simultaneous, five-dimensional fit. In addition to $M_{\rm bc}$, $\Delta E$, and $\Delta t$, we fit two angles in the transversity basis to measure the polarization and to statistically separate the $CP$ eigenstates.
We define the transversity basis with the $x$-axis pointing in the direction of the momentum of the $D^{*+}$ meson in the CMS, and the decay products of the $D^{*-}$ meson lying in the $xy$-plane in the rest frame of the $D^{*+}$ meson. The angle $\theta_{\rm tr}$ is defined as the angle between the momentum of the pion from the $D^{*+}$ decay and the $z$-axis in the $D^{*+}$ rest frame, and $\theta_1$ as the angle between the momentum of the pion from the $D^{*-}$ decay and the $x$-axis in the rest frame of the $D^{*-}$. The angular distribution of the $B^0$ decay products is given by
\begin{equation} 
\frac{1}{\Gamma}\frac{d^2\Gamma(B^0\rightarrow D^{*+}D^{*-})}{d\cos \theta_{\rm tr} d\cos \theta_1 } = \frac{9}{16} \displaystyle\sum_{i=0,\perp,\shortparallel} \hspace{-1pt} R_i H_i(\cos \theta_{\rm tr}, \cos \theta_1), \label{equ:ang}
\end{equation}
where
\begin{flalign}
  H_0 &= 2 \cos^2 \theta_1 \sin^2 \theta_{\rm tr}, \\
  H_\perp &= 2 \sin^2 \theta_1 \cos^2 \theta_{\rm tr},\\
  H_\shortparallel &= \sin^2 \theta_1 \sin^2 \theta_{\rm tr}
\end{flalign}
and $R_0$, $R_\perp$, and $R_\shortparallel$ are real-valued parameters that are defined to have a sum of one~\cite{theo_ang}. The term containing $R_\perp$ represents the $CP$-odd component, while the other two represent the $CP$-even components. The measured distributions differ from this theoretical expectation due to mis-reconstruction, the angular resolution of the slow pions, and varying reconstruction efficiency. The shapes of the three components are obtained separately from simulated signal events. The signal distributions are described by fourth-order polynomials, where the terms of odd order are fixed to zero in $\cos \theta_{\rm tr}$. The parameters $R_i$ are corrected for the relative reconstruction efficiencies of the different components.
The background in $\cos \theta_{\rm tr}$ is described by a polynomial of second order with the first order fixed to zero, and in $\cos \theta_1$ with a polynomial of fourth order.

\par
The decay vertex of the accompanying $B$ meson (referred to as the tag $B$) is obtained from charged tracks that are not used for the reconstruction of the signal $B$ meson. This is done by constraining them with a kinematic fit to originate from a common vertex and discarding tracks that likely originate from secondary decays. The flight length difference is given by $\Delta z = z_{CP} - z_{\rm tag}$. The procedure for flavor tagging is described in Ref.~\cite{TaggingNIM}. For each candidate, it gives the flavor $q$ of the tag $B$ meson and a tagging quality variable $r$ ranging from $r = 0$ for no flavor discrimination to $r = 1$ for unambiguous flavor assignment. The event sample is divided into seven bins of $r$. The wrong tag fraction $w$ and the wrong tag fraction difference between the tagging of $B^0$ and $\bar{B}^0$ mesons $\Delta w$ were determined in each of the bins using high statistics control samples~\cite{jpsiks}. Equation~\ref{equ:dt_sig} is modified to account for the influence of imperfect flavor tagging and the dilution by different $CP$-modes in the final state to
\begin{multline}
 \mathcal{P}_{B^0} \left( \Delta t, \cos \theta_{\rm tr}, \cos \theta_1 \right) = \frac{1}{4 \tau_{B^0}} e^{{}^{ -|\Delta t| } / {}_{ \tau_{B^0}}} \\ \times \bm{(} 1 - q \Delta w + q(1 - 2 w) \{ [1 - 2 P_{\rm odd}(\cos \theta_{\rm tr}, \cos \theta_1)]  \\ \times S \sin (\Delta m \Delta t) + A \cos (\Delta m \Delta t) \} \bm{)} .
\end{multline}
This modification includes the assumption that $S_{+} = -S_{-} = S$ and $A_+ = A_{-} = A$, where $S_+$ and $A_+$ describe the $CP$ violation in the $CP$-even and $S_-$ and $A_-$ in the $CP$-odd component. $P_{\rm odd}$ is the probability of an event to be $CP$-odd, derived from the angular distributions and the parameters $R_i$ with
\begin{equation}
 P_{\rm odd}(\cos \theta_{\rm tr}, \cos \theta_1) = \frac{R_\perp H_\perp(\cos \theta_{\rm tr}, \cos \theta_1)}{\displaystyle\sum_{i=0,\perp,\shortparallel} R_i H_i(\cos \theta_{\rm tr}, \cos \theta_1)}.
\end{equation}
The values of $\Delta m$ and $\tau_{B^0}$ are fixed to the current world averages of $507 \, \rm{ps}^{-1}$ and $1.525 \, \rm{ps}$, respectively~\cite{PDG}. This signal PDF is convolved with a term describing the detector resolution and effects from secondary decays of the tag $B$. The resolution function and its parameters were obtained from high statistics control samples and are described in detail in Ref.~\cite{jpsiks}. We multiply the resulting PDF with Eq. \ref{equ:ang} and the according PDF in $M_{\rm bc}$ and $\Delta E$ to obtain the overal signal PDF.
The background in $\Delta t$ is modeled with the sum of an exponential and a prompt component, convolved with the sum of two Gaussian distributions whose widths depend on the uncertainty of $\Delta z$.

\par
\begin{figure}[ht]
 \includegraphics[width=.45\textwidth]{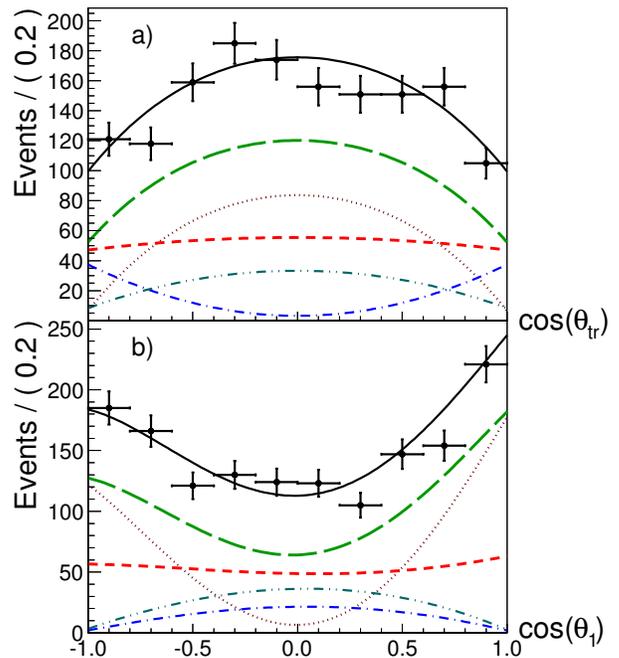}
\caption{Fit projections and data points of $\cos \theta_{\rm tr}$ (top) and $\cos \theta_1$ (bottom) in the signal region of $M_{\rm bc}$ and $\Delta E$. The black, solid line shows the total fitted function, the green, long dashed line the fitted signal, and the red, short dashed line the background. The brown dotted, the blue dash dotted, and the cyan dash double dotted lines show the contributions of the $R_0$, $R_\perp$, and $R_\shortparallel$ components, respectively.}
\label{fig:ang}
\end{figure}
\begin{figure}[ht]
 \includegraphics[width=.45\textwidth]{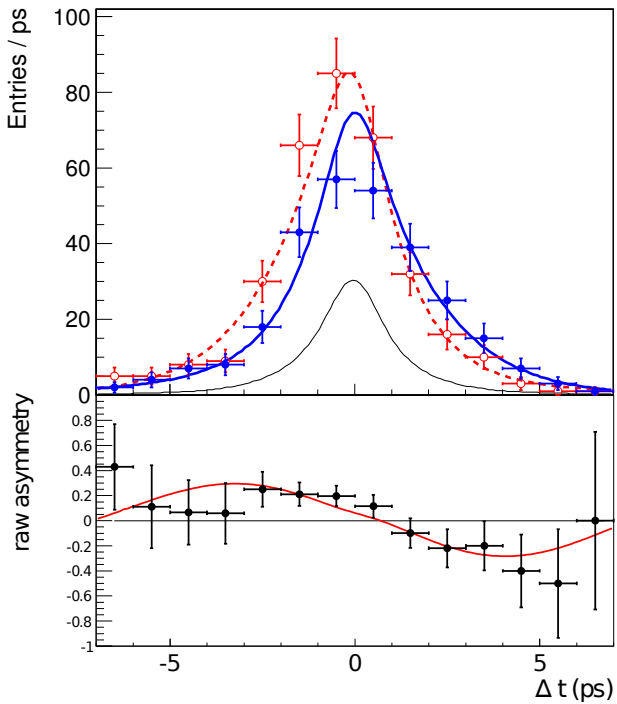}
\caption{The top plot shows events with $r > 0.5$ for $q = 1$ and $q = -1$ in the signal region of $M_{\rm bc}$ and $\Delta E$. The red, dashed (blue, solid) line and markers show the data points and the corresponding fit of events for $q = 1 \, (q = -1)$. The thin, black line shows the estimated background contribution. The bottom plot shows the raw asymmetry of the two top curves.}
\label{fig:dt}
\end{figure}

The fraction of signal events is determined separately for each bin in $r$ using the fit model in $\Delta E$ and $M_{\rm bc}$, which is the same as that used in the branching fraction measurement. In total, the PDF describing the signal and background in $M_{\rm bc}$, $\Delta E$, $\cos \theta_{\rm tr}$, $\cos \theta_1$, and $\Delta t$ has 30 free parameters, with four of them being $S$, $A$, $R_0$, and $R_\perp$. They are determined with an unbinned maximum likelihood fit. We find
$R_0 = 0.624 \pm 0.029, R_\perp = 0.138 \pm 0.024, S = -0.79 \pm 0.13$, and $A = 0.15 \pm 0.08,$
with the statistical correlations given in Table \ref{tab:corr}. Figure \ref{fig:ang} shows the fit projections and the data points for the two angles. Figure \ref{fig:dt} shows the $\Delta t$ distribution of well-tagged events ($r > 0.5$) with $q = 1$ and $q = -1$ and the corresponding raw asymmetry $\left(N_{+}(\Delta t) - N_-(\Delta t) \right)/\left(N_{+}(\Delta t) + N_-(\Delta t) \right)$, with $N_i (\Delta t)$ being the number of events with flavor tag $i$ in decay time bin $\Delta t$.

\begin{table}[ht]
\caption{Statistical correlations of the physical parameters.}
\label{tab:corr}
\begin{tabular}
 {c c c c}
\hline \hline
          &  $A$       & $R_0$   & $R_\perp$ \\ \hline
$S$       &  $0.07 $   & $0.01 $ & $-0.17 $  \\
$A$       &            & $-0.01$ & $-0.01 $  \\
$R_0$     &            &         & $-0.14 $  \\
\hline \hline
\end{tabular}
\end{table}

\par
The systematic errors are summarized in Table \ref{tab:sys_cp}. Major contributions come from the signal resolution function, the vertex reconstruction, and tag side interference. The errors due to the uncertainties of the vertex resolution function are studied by varying each fixed parameter obtained from data (simulated events) by one (two) standard deviations and repeating the fit. The interference of Cabibbo-favored $b \to c \bar{u} d$ and doubly Cabibbo-suppressed $\bar{b} \to \bar{u} c \bar{d}$ decays of the tag $B$ meson is noted as tag side interference and described in detail in Ref.~\cite{TSI}. Its influence is estimated by generating random data samples from a modified PDF that includes tag side interference and fitting them with the nominal PDF. The difference between the fit result and the input value is taken as the systematic error. The vertexing error combines multiple sources of error such as the requirement for the vertex fitting quality variable, the selection in $\Delta t$, and the selection criteria for the tracks used for the determination of the decay vertex of the tag $B$. It also contains systematic errors due to imperfect SVD alignment and potential bias in the measurement of $\Delta z$, which are estimated with Monte Carlo simulations. The fit model error has been estimated in the same manner as done for the branching fraction measurement. The contributions due to the physics parameters, the flavor tagging model, and the reconstruction efficiencies of the three polarizations are all estimated by varying the corresponding values within their uncertainties. The total systematic error is calculated by adding all contributions in quadrature.
The total systematic error on $S$ is reduced significantly in comparison to the previous Belle measurement by including $R_\perp$ as a free parameter in the fit for $S$ and $A$.

\begin{table}[ht]
\caption{Systematic errors of $S$, $A$, $R_0$, and $R_\perp$.}
\label{tab:sys_cp}
\begin{tabular}
{lc@{\hspace{0.10cm}}c@{\hspace{0.10cm}}c@{\hspace{0.10cm}}c}
\hline \hline
 & $S$ & $A$ & $R_0$ & $R_\perp$ \\ \hline
Fit model &                       $\pm 0.002$ & $< 0.001$   & $\pm 0.010$ & $\pm 0.003$ \\
Physics parameters &              $\pm 0.004$ & $\pm 0.001$ & $\pm 0.001$ & $< 0.001$   \\
Flavor tagging &                  $\pm 0.003$ & $\pm 0.002$ & $< 0.001$   & $< 0.001$   \\
Tag side interference &           $\pm 0.007$ & $\pm 0.032$ & $\pm 0.002$ & $\pm 0.001$ \\
$\Delta t$ signal resolution &    $\pm 0.021$ & $\pm 0.006$ & $\pm 0.001$ & $\pm 0.001$ \\
Reconstruction efficiencies &     $< 0.001$   & $< 0.001$   & $\pm 0.002$ & $\pm 0.001$ \\
Vertexing &                       $\pm 0.017$ & $\pm 0.021$ & $\pm 0.004$ & $\pm 0.004$ \\ \hline
Total & 			  $\pm 0.029$ & $\pm 0.038$ & $\pm 0.011$ & $\pm 0.006$ \\
\hline \hline
\end{tabular}
\end{table}

\par
We estimate the significance of the fit according to Wilks' theorem~\cite{wilks}. The hypothesis of no $CP$ violation $(S=A=0)$ is excluded with a significance of $5.4 \, \sigma$. This significance takes into account the systematic uncertainty by convolving the likelihood function with a Gaussian distribution whose width is equal to the total additive systematic error. The measurement is consistent with the Standard Model expectation of $A \approx 0$ and $S = \sin 2 \phi_1$~\cite{Xing} within $1.4 \, \sigma$.

\par
Additionally, we release the assumption, that $S_{+} = -S_{-}$ and $A_{+} = A_{-}$. This allows for the possibility of different relative penguin contributions to the two different components in the fit. The according signal PDF in $\Delta t$ is given by
\begin{multline}
 \mathcal{P}_{B^0} ( \Delta t) = \frac{1}{4 \tau_{B^0}} e^{{}^{ -|\Delta t| } / {}_{ \tau_{B^0}}} \bm{(} 1 - q \Delta w + q(1 - 2 w) \\ 
\times \{ [(1 - P_{\rm odd}) S_+ + P_{\rm odd} S_- ] \sin (\Delta m \Delta t) \\ 
 + [(1 - P_{\rm odd}) A_+ + P_{\rm odd} A_- ] \cos (\Delta m \Delta t) \} \bm{)}.
\end{multline}
Note the difference in the sign convention of $S_-$ in comparison to $S_\perp$ in Ref.~\cite{DC_BaBar}.
We obtain $S_+ = -0.81 \pm 0.13 \pm 0.03$, $A_+ = 0.18 \pm 0.10 \pm 0.05$, $S_- = 1.52 \pm 0.62 \pm 0.12$, and $A_- = -0.05 \pm 0.39 \pm 0.08$. Within the uncertainties we find no indication for different $CP$ violation in the $CP$-even and $CP$-odd component.

\par
In summary, we report the measurement of the branching fraction, the polarization, and the parameters of time-dependent $CP$ violation of $B^0 \rightarrow D^{*+} D^{*-}$ decays. The results are 
\begin{align*}
&\mathcal{B}_{(B^0 \rightarrow D^{*+} D^{*-})} = \left[ 7.82 \pm 0.38\rm{(stat.)} \pm 0.63\rm{(syst.)} \right] \times 10^{-4},\\
&R_0 = 0.624 \pm 0.029\rm(stat.) \pm 0.011\rm{(syst.)},\\
&R_\perp = 0.138 \pm 0.024\rm{(stat.)} \pm 0.006\rm{(syst.)},\\
&S = -0.79 \pm 0.13\rm{(stat.)} \pm 0.03\rm{(syst.)}, \text{and}\\
&A = 0.15 \pm 0.08\rm{(stat.)} \pm 0.04\rm{(syst.)}.
\end{align*}
These are consistent with previous measurements by the Belle collaboration~\cite{BRF_Belle, DC_Belle2} and supersede them. It is also in agreement with the Standard Model expectation and is the first measurement of $B^0 \rightarrow D^{*+} D^{*-}$ decays that exhibits $CP$ violation with a significance greater than $5 \, \sigma$. 

% %***** Acknowledgments *****
\par
We thank the KEKB group for the excellent operation of the accelerator; the KEK cryogenics group for the efficient operation of the solenoid; and the KEK computer group, the National Institute of Informatics, and the PNNL/EMSL computing group for valuable computing and SINET4 network support.  We acknowledge support from the Ministry of Education, Culture, Sports, Science, and Technology (MEXT) of Japan, the Japan Society for the Promotion of Science (JSPS), and the Tau-Lepton Physics Research Center of Nagoya University; the Australian Research Council and the Australian Department of Industry, Innovation, Science and Research; the National Natural Science Foundation of China under contract No.~10575109, 10775142, 10875115 and 10825524; the Ministry of Education, Youth and Sports of the Czech Republic under contract No.~LA10033 and MSM0021620859; the Department of Science and Technology of India; the Istituto Nazionale di Fisica Nucleare of Italy; the BK21 and WCU program of the Ministry Education Science and Technology, National Research Foundation of Korea, and GSDC of the Korea Institute of Science and Technology Information; the Polish Ministry of Science and Higher Education; the Ministry of Education and Science of the Russian Federation and the Russian Federal Agency for Atomic Energy; the Slovenian Research Agency;  the Swiss National Science Foundation; the National Science Council and the Ministry of Education of Taiwan; and the U.S.\ Department of Energy and the National Science Foundation. This work is supported by a Grant-in-Aid from MEXT for Science Research in a Priority Area (``New Development of Flavor Physics''), and from JSPS for Creative Scientific Research (``Evolution of Tau-lepton Physics'').


\begin{thebibliography}{99}
\bibitem{CC}
Throughout this paper, the inclusion of the charge-conjugate decay mode is implied unless otherwise stated.
\end{thebibliography}
\end{document}